\documentstyle[prl,aps,multicol,psfig]{revtex}

\begin{document}

\title{ Temporal Scaling of
                          Interfaces Propagating in Porous Media}

\author{ Viktor K. Horv\'ath\cite{fn0}   and H. Eugene Stanley}

\address{ Center for Polymer Studies and Department of Physics, \\
          Boston University, Boston, Massachusetts 02215}

\date{Physical Review E, pp.5166-5169, {\bf 52} 1995 Nov.}

\maketitle

\begin{abstract}
To better understand the temporal behavior of a roughening meniscus
driven by capillary forces during the imbibition of a viscous fluid in
porous media, we measure the height-height autocorrelation function
$C(t)$ using a constant driving force.  We find $C(t)\sim t^{\beta}$,
with $\beta = 0.56\pm 0.03$, and provide the first experimental evidence
for driving force independent temporal scaling behavior of a propagating
wetting front in the presence of quenched noise.  We interpret the value
of ${\beta}$ in terms of the possibility that the dynamics may be
governed by nonvanishing non-linearity due to anisotropic depinning.
\end{abstract}


\begin{multicols}{2}

The growth of rough surfaces and interfaces under far-from-equilibrium
conditions is a common phenomenon in nature \cite{General}.  Examples
include such processes as vapor deposition, crystallization, thin film
growth by atomic beams, settling of granular materials, and fluid flow
in porous media.  The fluctuations of the interface height $h(x,t)$ can
be characterized either by their standard deviation $\sigma(l,t)$, or by the
height-height correlation function
$C(l,t)\equiv[\langle ( {\tilde h}(l+x,t+\tau) - {\tilde
h}(x,\tau))^2\rangle_{x,\tau}]^{1/2}$, where
${\tilde h}(x,\tau)\equiv h(x,\tau)-\langle h(x,\tau)\rangle_x$.
The common belief is that $\sigma(l,t)$ and $C(l,t)$ exhibit the same
general statistical properties.


Particularly useful is the Family-Vicsek dynamic scaling hypothesis
\cite{FV:dynamic},
\begin{equation}
C(t)=C(0,t)\sim L^{\alpha}{\cal F}(t/L^{\alpha /\beta}),
\label{FV:scaling}
\end{equation}
where $L$ is the system size and ${\cal F}(y)\sim y^\beta$ for $y\ll 1$,
${\cal F}(y)\sim {\text const}$ for $y\gg 1$.
Extensive numerical simulations of different computer models
\cite{General} predict universal values for the exponents $\beta$ and
$\alpha$.  Theoretical work is based on Langevin-type equations, such as
\begin{equation}
{\partial {h({\vec {\bf x}},t)}\over\partial t}=v_0 +
\nu\nabla^2h+\lambda_{eff}(\nabla h)^2 + \eta({\vec {\bf x}},t).
\label{eq:Langevin}
\end{equation}
The linear ($\lambda_{eff}=0$) and the non-linear ($\lambda_{eff}
\not=0$) cases are referred to as the Edwards-Wilkinson (EW) \cite{EW}
and Kardar-Parisi-Zhang (KPZ) \cite{KPZ} equations.  Analytical solution
of Eq.~(\ref{eq:Langevin}) in $d'=1$ (where $d'$ is the dimension of
${\vec {\bf x}}$) reveals the {\it same\/} spatial exponent
($\alpha=1/2$), but {\it different} temporal exponents
\begin{mathletters}
\label{Bexponents}
\begin{equation}
{\beta}_{EW}=1/4, ~~{\beta}_{KPZ}=1/3.
\label{solution:Tbeta}
\end{equation}
If we know these exponents, then we can distinguish between linear and
non-linear dynamics based on $\beta$. Therefore it is important to make
accurate measurements of $\beta$.


If the randomness in the dynamics is due to the inhomogeneity of the
media where the moving phase is propagating, then the resulting
interface exhibits different scaling behavior.  This time-{\it
independent} ``quenched'' randomness can be described by replacing
$\eta({\vec {\bf x}},t)$ by $\eta({\vec {\bf x}},h(t))$ in
Eq.~(\ref{eq:Langevin}).  We denote the linear and non-linear equations
as QEW and QKPZ respectively.
Numerical solution of the QEW equation \cite{QEW}
produces self--affine interfaces with a roughness exponent $\alpha$ in
the range $[0.5-1.0]$, that is tunable with the driving force $v_0$
\cite{KESSLER:alpha,Percolation:models,PARISI:beta,CSAHOK:dimanal,Robbins:NewPRL,Kim,Kertesz:Wolf,Lopez:Euro}.
Different computer models also exhibit very scattered results for
$\alpha$ \cite{Percolation:models}.  However, theoretical considerations
suggest \cite{PARISI:beta} that $\beta_{QEW} =(4-d')/4$.  Numerical
integration of the QKPZ equation close to the pinning transition
exhibits temporal scaling with $\beta=0.61\pm0.06$, in good agreement
with the result of simple dimensional analysis which predicts
$\beta_{QKPZ} =(4-d')/(4+d')$ \cite{CSAHOK:dimanal}. Hence for $d'=1$:
\begin{equation}
\beta_{QEW}=3/4, ~~{\beta}_{QKPZ}=3/5.
\label{solution:Qbeta}
\end{equation}
The fact that scaling of surfaces during roughening in the presence of
quenched noise exhibits driving force independent temporal scaling also
underlines the importance of the investigation of the exponent $\beta$,
which is the main goal of this Letter.
\end{mathletters}


\begin{figure}[h]
  \begin{center}
    \leavevmode
     \psfig{figure=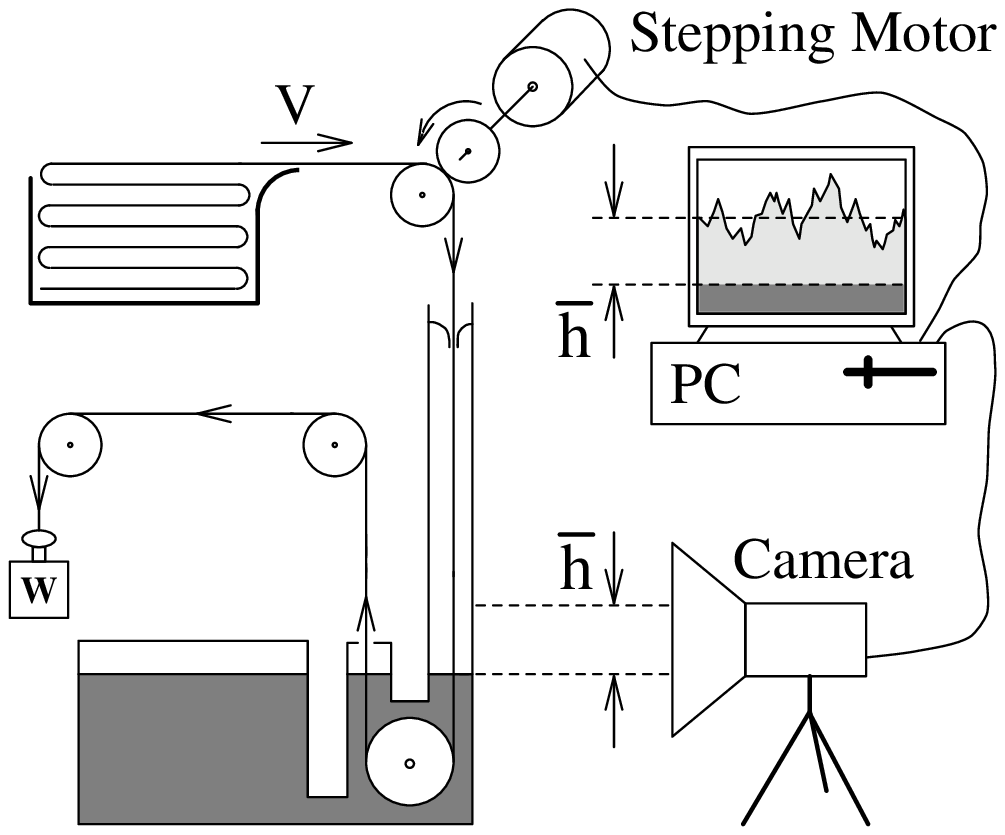,clip=true,width=8cm}
  \end{center}
{\ FIG. 1. }
Schematic of the experimental setup.  The average height ${\overline h}$
of the meniscus in the wetted paper was kept at a constant level. A video
camera digitized the interface in real time and the computer pulled down
the paper if $\overline h$ became larger than the predefined value $H$.
\end{figure}
Specifically, we investigate the growing interface during imbibition of
viscous liquids in filter paper.  The experimental setup (Fig.~1)
contains two vertically-positioned 40cm$\times$30cm parallel plexiglass
plates separated from each other by 3 cm and closed with side walls.
The bottom part of the plates extend into a liquid container from which
water starts to wet the filter paper.  We designed the size of this
liquid container to be large enough to keep the free surface of water at
a constant level.  All height measurements are relative to this
reference level.  The upper part of the cell is closed by polyethylene
film to prevent a large amount of evaporation.  The dry paper passes
through two touching cylinders driven by a stepping motor, and enters
into the cell through a narrow gap on this film.  We use other rollers
and a proper weight to keep the paper stretched.
Stretching of wet paper sometimes involves changes in the paper structure
mostly by elongating the paper strip.
This was checked by double
measurements of the speed of the paper strip. We measured the speed
of the paper both at the driving rollers at the stepping motor,
and at the weight used for stretching the paper. Any dilatation of the
paper strip would appear in a difference between the speeds at these points.
We have used enough small stretching weight to avoid any such a difference.


We monitor the wetting front $h(x,t)$ with a high resolution (0.48
million pixels) CCD NTSC camera horizontally centered in front of the
front plate.  We use a personal computer to digitize the video image,
and to calculate the average height ${\overline h}=\langle
h(x,t)\rangle_x$ in {\it real time}.  We set our video system to
digitize a 8.4\/cm wide segment of the filter paper strip involving a
$110\mu$m (squared) pixel size.  We control the motion of the paper by a
gear mechanism driven by a stepping motor connected to the same
computer.  According to our calibration, one step of the motor
corresponds to a paper shift of $\delta y=85\mu$m.


The water front moves upward in the paper due to the
{\it effective} driving force
$\varepsilon$, which is mainly determined by the balance between
capillarity and gravity.  We maintain this driving force constant by
holding ${\overline h}$ at a predefined value $H$.  If ${\overline h}$
exceeds $H$ by $\delta y$, then we pull the paper down by $\delta y$.
This negative feedback, apart from fluctuations, prescribes a constant
average speed $V$ for the paper for a given $H$.  Our measurements at 14
different values of $H$ fit remarkably well the power law
\begin{equation}
V\propto H^{-\Omega}, ~~~~~
\label{scaling:H_V}
\end{equation}
where $\Omega=1.594\pm 0.007$, and the quoted uncertainty is the
standard deviation from the fitted value.  The relation
(\ref{scaling:H_V}) holds in the entire investigated range of the
control parameter, $H \approx$~[2mm,40mm], which spans approximately 2.5
decades of the capillary number.


\begin{figure}[h]
  \begin{center}
    \leavevmode
     \psfig{figure=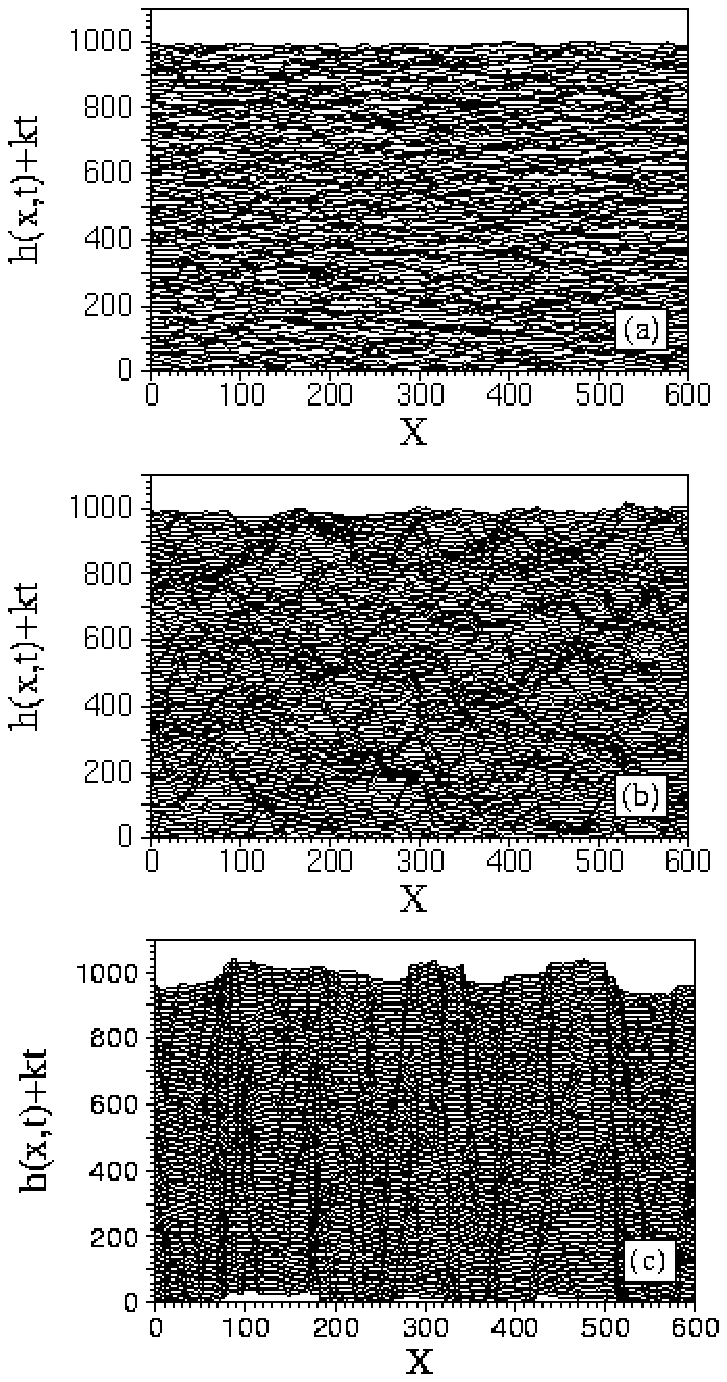,clip=true,width=8cm}
  \end{center}
{\ FIG. 2. }
Snapshots of the evolving interface for different velocities $V$. 
Units are in pixels.
(a) $V=4.03\times 10^{-2}$cm/sec, (b) $V=1.32\times 10^{-2}$cm/sec, (c)
$V=1.40\times 10^{-3}$cm/sec.  The sampling rate is the same constant
for all pictures and the artificial shift $k$ between the contour lines
is proportional to this constant.
\end{figure}
During the experiment, the computer saves the contour lines of the
digitized images for later analysis.  Snapshots of the evolving
interface at different times are superposed in Fig. 2 for three
different values of the velocity.  We find only a few and negligible
\cite{Robbins:NewPRL} overhangs, and when they occur we remove them in
the usual way\cite{Mitsugu}.


\begin{figure}[h]
  \begin{center}
    \leavevmode
     \psfig{figure=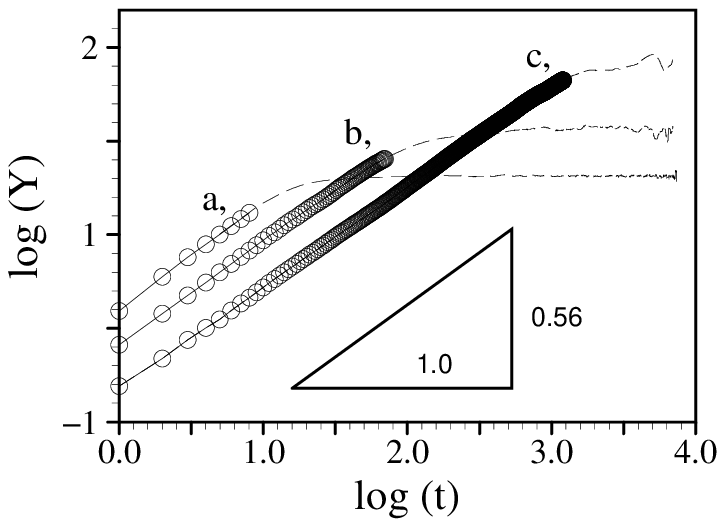,clip=true,width=8cm}
  \end{center}
{\ FIG. 3. }
The differential autocorrelation function $Y$ vs. time is shown by
circles.  The dashed lines present the saturation regime of the
autocorrelation function.  The speeds for the 3 curves are the same as
in Fig. 2. Larger speed implies larger intrinsic width (intercept), and
smaller saturation value.
\end{figure}
Using the contour lines we calculated $C(t)$, which has less uncertainty
than the standard deviation of the interface height, and is therefore
more useful for calculating the true exponents \cite{Kim}.  To check the
scaling behavior of $C(t)$, we form the quantity
\begin{equation}
Y^2(t_i)\equiv {{C^2(t_{i+1}) - C^2(t_i)}
\over{t_{i+1}^{2\beta'}-t_i^{2\beta'}}}\times {t_{i}^{2\beta'}}
\label{definition:ofY}
\end{equation}
where $t_i,t_{i+1}$ are successive times.  If $C(t) \propto t^{\beta}$,
then  $Y(t)|_{\beta'=\beta}$ must scale
the same way, but provides a more precise (self-consistent) method to
determine $\beta$ by eliminating  any existing
additive ``correction-to-scaling'' terms \cite{Kertesz:Wolf}. Figure 3
presents our results for the same data sets as shown in Fig.~2.  We find
$Y(t) \propto t^{\beta}$ for early times and this power-law behavior
extends over slightly more than three decades at the minimum driving
force (${V}=1.4\times 10^{-3}$cm/s).  Within the experimental
uncertainty, ${\beta}$ is independent of our control parameter $H$,
yielding an average value
\begin{equation}
\label{averagevalue}
\beta_{\exp}=0.56\pm 0.03.
\end{equation}


This result can be compared to
2 dimensional theoretical results and simulations because
the small thickness of the filter paper (which is in the order of microns)
does not permit any coarse-grained roughening
perpendicular to the paper strip.
Comparing ${\beta_{\exp}}$ to (\ref{solution:Tbeta}), we may exclude the
dominance of annealed randomness, which prescribes a much lower value for
$\beta$.  Moreover, the value of
$\beta_{\mbox{\scriptsize QEW}}$ is also inconsistent with our result.
Nevertheless quenched disorder with non-linear dynamics produces a
remarkably close exponent, $\beta_{\mbox{\scriptsize QKPZ}}=0.6$
\cite{CSAHOK:dimanal}, which suggests that
quenched randomness and non-linearity play important roles
in the dynamics of vertical
imbibition.

Remarkably, we find no crossover for $\beta$ in the scaling regime, as
might have been expected in the case of QEW
\cite{PARISI:beta,Lopez:Euro}, where the temporal scaling is described
by $\beta_{\mbox{\scriptsize QEW}}$ for early times and by
$\beta_{\mbox{\scriptsize EW}}$ for late times (at an intermediate
driving force). As the driving force increases, the crossover point
$t^*$ becomes smaller and the scaling region described by
$\beta_{\mbox{\scriptsize EW}}$ enlarges.  From Fig.~3, we
observe crossover neither as a function of driving force, nor as a
function of time---even at the at the maximum driving force
(${V}=2.1\times 10^{-1}$cm/s).  The absence of such a crossover also
indicates the importance of the non-linearity in our experimental
system.


It has been observed \cite{Amaral94b} that numerical results fall into
two groups depending on the origin of the non-linear term $\lambda
(\nabla h)^2$.  Kinematics produces a $\lambda$ which vanishes at the
threshold and the resulting interface belongs to the same university
class where the non-linear term is absent, $\lambda_{\mbox{\scriptsize
eff}}=0$.  On the other side the QKPZ university class is characterized
by a non-vanishing non-linear term leading to
($\lambda_{\mbox{\scriptsize eff}}\not =0$).
What is the origin of such a non-linearity which leads to
$\beta_{\exp}\approx\beta_{\mbox{\scriptsize QKPZ}}$ in our experiment?
Recently Tang,
Kardar, and Dhar demonstrated \cite{TKD} that {\it anisotropic
depinning} yields a non-zero $\lambda_{\mbox{\scriptsize eff}}$ at the
depinning transition.   To illustrate the existence of such an anisotropy in
our system, we consider the filter paper as an interconnected
network of different capillary tubes.
The inhomogeneity of this network can be considered as a
random field with amplitude $\Delta$.  This random field is
correlated isotropically in space within a distance $a$, but the driving
force (pressure drop) is different along horizontal and vertical tubes,
$F_h=f$ and $F_v=f-\rho ga$ respectively, because the driving force must
compensate the weight of the liquid in a vertical tube ($\rho$ is the
density of the liquid).  Due to this anisotropy,  a segment of the
interface can easier be pinned vertically than horizontally.  Therefore
a slope-dependent effective driving force
${\tilde F(\nabla h)}=F(\nabla h)-F_c$ is
generated under coarse graining, which explains the absence of large
overhangs.  Tang et al. pointed out \cite{TKD} that an
expansion of ${\tilde F(\nabla h)}$ around its minimum yields to a term
$\lambda_{\mbox{\scriptsize eff}}(\nabla h)^2$ which remains finite,
independent of $V$, which is the hallmark of the QKPZ universality class.

Figure 3 demonstrates two effects of increasing the
driving force, namely the intercepts increase and the saturated
values decrease.  In order to incorporate this behavior into the scaling
formalism, we assume that Eq.~(\ref{FV:scaling}) is valid with the
rescaled time and space variables ${t'}^\beta={t}^\beta\/V^{\theta_t}$
and ${L'}^\alpha={L}^\alpha\/V^{-\theta_L}$.  As a consequence of this
assumption, $C(t)_{L,V}$ scales as
\begin{equation}
C(t)_{\mbox{\scriptsize L,V}}\sim V^{-\theta_L}\/{L}^{\alpha}
{\cal F}\left({t\/L^{-\alpha/\beta}}\/V^{(\theta_t+\theta_L)/\beta}\right).
\label{EDS:F}
\end{equation}
This extended dynamic scaling (EDS) hypothesis is fully compatible with
Eq.~(\ref{FV:scaling}) for a given $V$, but also describes the
characteristic role of the driving force in presence of quenched
disorder.  In the limit of small driving force and early times
($t^{\beta}\/{V}^{\theta_t+\theta_L}\ll L^\alpha$) the EDS
hypothesis reduces to $C\sim t^{\beta}\/{V}^{\theta_t}$.  For $t\gg
t_c$, where $t_c$ is a critical time defined by
$t_c=L^{\alpha/\beta}{V}^{-(\theta_t+\theta_L)/\beta}$, the
autocorrelation function is independent of time and saturates at a
constant value $C\sim L^{\alpha}\/{V}^{-\theta_L}$.  This asymptotic
behavior is consistent with Fig.~3, as it is expected.
In order to
perform a quantitative check of Eq.~(\ref{EDS:F}), we plot
$Y'\equiv C\/L^{-\alpha}V^{\theta_L}$
vs.
$X'\equiv t\/L^{-\alpha/\beta}\/V^{(\theta_t+\theta_L)/\beta}$
for different values of
$\theta_t$ and $\theta_L$.  The best data collapse is found with
$\theta_t=0.37$ and $\theta_L=0.48$.  Using these exponents we form the
scaling plot for measurements at 5 different values of $H$ and times
spanning three decades.  From the data collapse in Fig.~4, we conclude
that our measurements are fully consistent  our scaling {\it Ansatz\/}.

\begin{figure}[h]
  \begin{center}
    \leavevmode
     \psfig{figure=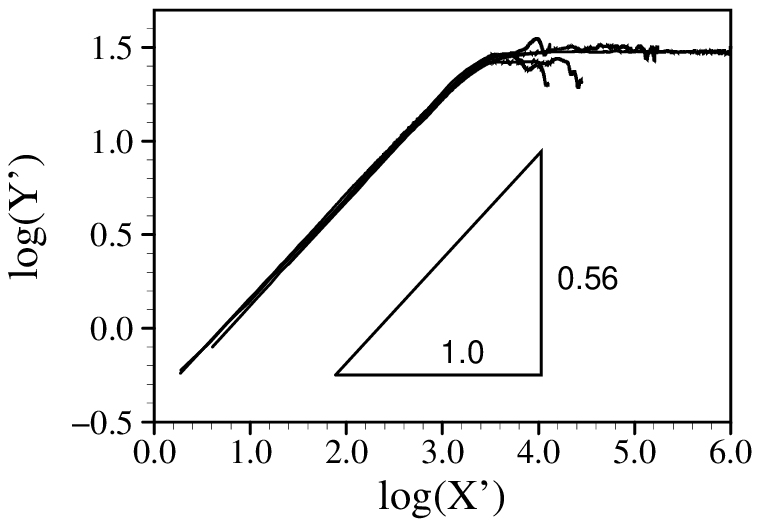,clip=true,width=8cm}
  \end{center}
{\ FIG. 4. }
  Scaling plot for five different data sets. In addition to the speeds
  of sets on Fig. 2 two additional sets are also drawn for $V=2.50\times
  10^{-3}$cm/sec) and $V=4.20\times 10^{-3}$cm/sec; $Y'$ and $X'$ are
  defined in the text.  The best result is achieved with exponents
  $\beta=0.56,~\theta_t=0.37,~\theta_L=0.48$.  According to the scaling
  function ${\cal F}$ the different data sets collapse on two straight
  lines, with slopes $\beta=0.56$ and zero for early and late times
  respectively.
\end{figure}

In fact there is some theoretical base for the EDS hypothesis: Kert\'esz
and Wolf \cite{KW} pointed out that a new diverging length
$\xi\sim|\varepsilon|^{-\nu}$ in the dynamics modifies the scaling of
the interface and $\xi$ must appear in the scaling form
$w(\varepsilon,L,t)\sim \xi^{{\alpha}'}{\cal H}(t',L')$, where $L'\equiv
L/\xi$, $t'\equiv t/{\xi}^{z'}$, and $z'\equiv {\alpha}'/{\beta}'$.  Far
from the transition point ${\cal H}(t',L')\sim {L'}^{\alpha}{\cal
F}({t'}/{L'}^{\alpha /\beta})$, which involves $w(\varepsilon\not
=0,L,t)\sim \xi^{{\alpha}'-\alpha}\/L^\alpha{\cal
F}(t/L^z\/\xi^{z-z'})$.  This expression resembles Eq.~(\ref{EDS:F}),
and the similarity is more than merely formal.  Movement of the meniscus
consists of pinning and depinning processes\cite{Oligami}.  Computer
simulations demonstrate \cite{Percolation:models} that as the driving
force $\varepsilon$ decreases, the surface slows down according to
$V\sim\varepsilon^{\theta}$ and the linear size of the pinned regimes
$\xi$ diverges $\xi\sim\varepsilon^{-\nu}$.  These two relations can be
integrated, with $\xi\sim V^{-\nu/\theta}$.  According to this picture,
we also expect a new diverging length in our system.  The relation
${\overline h}\sim V^{-1/\Omega}$ points in this direction, although it
is clear that Eq.~(\ref{scaling:H_V}) must break down at the pinning
transition, where $V=0$.  If we now replace $\xi$ by $V^{-\nu/\theta}$
in $w(\varepsilon\not =0,L,t)$, and compare the result to
Eq.~(\ref{EDS:F}), we see that there is a mapping between the two
relations with ${\alpha}'=\alpha+\theta\theta_L/\nu$ and
$z'=z+\theta(\theta_t+\theta_L)/(\beta\nu)$.  Kert\'esz and Wolf
demonstrated \cite{KW} that ${\alpha}'$ and $z'$ are the characteristic
exponents at the transition point between two different morphological
phases. It remains an interesting open question if these exponents have
the same significance at the pinning transition too, and this question
requires more experimental investigation.


To summarize, we have studied the temporal behavior of roughening
interfaces during vertical imbibition in porous media.  The
height-height autocorrelation function $C(t)_{L,V}$ of the meniscus
exhibits temporal scaling without crossover.  The corresponding exponent
is {\it independent} of driving force, $\beta=0.56\pm 0.03$.  We
conclude that nonlinearity plays important role in the dynamics of our
system.  For a compact description of our measurements, we suggest an
extended dynamic scaling hypothesis, which describes the characteristic
role of the driving force in the presence of quenched disorder, and is
fully compatible with the dynamic scaling developed for systems with
annealed randomness.  From the resulting data collapse, we find
$\theta_t=0.37$, $\theta_L=0.48$.  Finally, we note that Rubio et
al. \cite{RUBIO:exp} measured the interface width $\sigma(l,t)$ versus
$l$ in a different experiment, and they reported a velocity dependent
scaling $w(l)\sim l^\alpha\/v^{-0.47}$, a result consistent with our
value of $\theta_L$.  Although our result suggests that $\beta$ is
independent of the dynamics of the system, it remains to be seen whether
the moments of the quenched noise play an important role in determining
the universality of exponents.



\vfill
\end{multicols} 
\end{document}